# Community Shaping in the Digital Age: A Temporal Fusion Framework for Analyzing Discourse Fragmentation in Online Social Networks

Amirhossein Dezhboro [1], Jose Emmanuel Ramirez-Marquez [1*], Aleksandra Krstikj [2]

[1*]School of Systems and Enterprises, Stevens Institute of Technology, 525 River St, Hoboken, 07030, NJ, USA.

[2]Escuela de Arquitectura, Arte y Diseño, Tecnologico de Monterrey, 52926 Lopez Mateos, Estado de Mexico, Mexico.

*Corresponding author(s). E-mail(s): jmarquez@stevens.edu;
Contributing authors: adezhbor@stevens.edu;
sandra.krstik@tec.mx;

**Abstract**

This research presents a framework for analyzing the dynamics of online communities in social media platforms, utilizing a temporal fusion of text and network data. By combining text classification and dynamic social network analysis, we uncover mechanisms driving community formation and evolution, revealing the influence of real-world events. We introduced fourteen key elements based on social science theories to evaluate social media dynamics, validating our framework through a case study of Twitter data during major U.S. events in 2020. Our analysis centers on discrimination discourse, identifying sexism, racism, xenophobia, ableism, homophobia, and religious intolerance as main fragments. Results demonstrate rapid community emergence and dissolution cycles representative of discourse fragments. We reveal how real-world circumstances impact discourse dominance and how social media contributes to echo chamber formation and societal polarization. Our comprehensive approach provides insights into discourse fragmentation, opinion dynamics, and structural aspects of online communities, offering a methodology for understanding the complex interplay between online interactions and societal trends.

1. **Introduction**

While the term "Community" is often linked to a group sharing a geographical space or common trait, it more broadly signifies any unified group with shared values or experiences, irrespective of physical proximity. Historically, locality heavily influenced community formation, resulting in traditional, location-based units like countries or cities (Gallagher and Savage 2013). However, the rise of the internet, especially social media, has transcended geographic boundaries, allowing the formation of diverse online communities based on shared interests.

The effects of this digital interconnectedness are profound. Research has shown that social media platforms can be instrumental in shaping individual identity and self-discovery, particularly for marginalized groups (Mandel 2019; Villa-Nicholas 2019). For example, LGBTQ+ students have found validation and connection through online platforms (Sallafranque-St-Louis and Normand 2017) and those on the autism spectrum have experienced reduced social isolation (Miller 2017)

However, this same anonymity and reach can be weaponized. Social networks have been used to mobilize and disseminate extremist ideologies (Sardarnia and Safizadeh 2019) Online spaces like alt-right communities can normalize violent rhetoric and fuel counter-hegemonic movements (Hodge and Hallgrimsdottir 2020). Understanding the mechanisms behind the formation and evolution of online communities is crucial for policymakers to mitigate potential risks while harnessing the positive potential of these digital spaces.

Building on the profound implications of online communities in shaping societal discourse, our research delves into the algorithmic identification and analysis of these digital assemblies. By leveraging advancements in machine learning and social network theory, we aim to unravel the intricate web of interactions and shared interests that bind individuals across the vast expanse of cyberspace. This study not only underscores the importance of understanding the dynamics of online communities for both social scientists and policymakers but also provides a methodological blueprint for future research in this burgeoning field. Our approach's methodological rigor and analytical depth reflect recent advancements in computational social systems, as evidenced by related works in the field (Babvey et al. 2020). Besides the novel framework we propose, our approach incorporates several novel features:
- Not Over-reliance on Predefined Metrics: Instead of depending heavily on predefined metrics (e.g. centrality measures from social network analysis theories) and showing an abstract representation of what is happening in our complex system, we adopt a more naive but informative approach to uncover deeper insights.
- Granularity: Our method allows for a fine-grained visualization of community behaviors, capturing the nuances of interactions and engagement.
- Four-layer Generalizable Framework: We propose a robust framework consisting of four layers that can be generalized across different social media platforms and community types.
- Visualization Formation with Information Fusion: We combine various data sources and visualization techniques to create comprehensive visual representations of community dynamics.

In this paper, we propose a new analysis framework based on thematic analysis (understanding the content and themes of discussions), interaction dynamics (focusing on the structure of user interactions and information flow), opinion dynamics (involving the formation, evolution, and polarization of opinions), and structural insights (focusing on the overarching characteristics of social media discussions and their broader implications). This framework captures several critical aspects:
- Users' engagement with topics
- Users' engagement with influential actors
- Users' reactions to topics
- Birth, growth, and death of dominant topics in social media
- The ratio of changes (engagement or reaction or growth or decline of topics in social media)
- Cohesiveness of topics
- Polarization (how divergent views become more extreme as like-minded individuals reinforce each other's beliefs)
- Segmentation (formation of different groups around specific topics)

Our research introduces a novel framework for identifying online communities on social media platforms. This framework leverages machine learning classification models to analyze user posts and interactions, revealing underlying group structures. These groups are then visualized through social network analysis, providing insights into their composition and dynamics. The innovative integration of text classification and social network analysis in this framework represents a significant contribution to the field, enabling more effective identification and understanding of online communities. Additionally, we have developed a temporal visualization method to represent the changes in these communities over time, providing a dynamic view of their evolution. We introduce fourteen key elements derived from the structure and dynamics of online social media, which encapsulate the insights expected from analyzing social structures and dynamics.

These elements are used to analyze the results of combining text and network data. Furthermore, we developed a case study using real-life data to demonstrate the practical application of our framework in analyzing the formation and shaping of online communities.

## 2. Literature Review

### 2.1. Topic Classification

Within the field of Natural Language Processing (NLP), a core task lies in text classification – the process of assigning predefined labels to textual data, enabling machines to interpret and organize unstructured information. This process has become increasingly vital with the growth of social media content, which presents unique challenges due to its informal language, brevity, and context-specific nuances. Recent studies have focused on enhancing classification accuracy by adapting models to understand social media text's informal nature better. Jotikabukkana et al. proposed an improved trained text model specifically for social media content, which shows promising results in accurately classifying such texts (Jotikabukkana et al. 2016). Sentiment analysis in social media has evolved from simple opinion mining to complex applications that can gauge public sentiment on a broad spectrum of topics. Applying deep neural networks to sentiment analysis has significantly improved the ability to understand and interpret the vast and varied data generated on social platforms. This advancement allows for a more nuanced understanding of public sentiment, enabling applications ranging from market analysis to political sentiment tracking. Moreover, integrating sentiment analysis with systematic literature reviews in social media has provided a new dimension to understanding public opinion and its evolution over time. This integration allows researchers to sift through massive datasets to identify prevailing sentiments and their shifts, offering invaluable insights into public perception and reaction to various stimuli (Dawei et al. 2021). Notably, the evolution from basic models to sophisticated architectures like Convolutional Neural Networks (CNNs) and Long Short-Term Memory (LSTM) networks has been instrumental in addressing complex linguistic structures and context-dependent interpretations inherent in textual data (Ahn, Plaisant, and Shneiderman 2014). In recent years, deep learning models have emerged as the preferred automated methods for text classification, owing to their exceptional performance and the elimination of complex manual feature extraction. These models leverage embedding techniques to map text into compact, continuous feature vectors, which are learned through neural networks (Minaee et al. 2020). Among the prominent contextual embedding models, ELMo (Peters et al. 2018) utilizes a three-layer bidirectional long short-term memory (LSTM) architecture. Following this, Google's BERT (Devlin et al. 2018), a bidirectional transformer model, gained widespread adoption for various natural language processing tasks, including text classification. These advancements underscore a paradigm shift towards models that encapsulate both semantic and syntactic nuances of language, thereby enhancing the granularity and accuracy of text classification.

### 2.2. Social Networks

The dynamics of social networks present a rich canvas for analyzing the multifaceted interactions and behaviors within digital communities. The advent of sophisticated visualization and analytical tools has significantly augmented our understanding of these networks, transcending beyond mere structural analysis to unravel temporal patterns and evolutionary trajectories. In this realm, the integration of temporal analysis with network visualization opens new avenues for examining organic growth, community formation, and the ebb and flow of influence amongst network entities. Such insights are invaluable for deciphering the underlying mechanisms driving social interactions and the dissemination of information across digital platforms. A crucial aspect of network analysis lies in the ability to visually represent the network's structure and identify key elements like hubs or communities (Kim and Hastak 2018).

The inherent social dynamics of Twitter have consistently attracted researchers seeking to analyze the connections and conversations that arise on the social media. The majority of these researches focus on the complex interplay between online discussions and real-world circumstances. For instance, Borelli et al. (Borrelli et al. 2022) examined how controversial offline circumstances can spark affective polarization - the extent to which opposing groups express dislike towards each other - in online social networks, and how counter-narratives by influential actors might mitigate this phenomenon. Li et al. (Li et al. 2012) proposed a framework for mining topic-specific influence in microblogging networks, aiding in the extraction and analysis of prevalent themes. Furthermore, Conover et al. (Conover et al. 2021) utilized social network analysis to delve into political polarization on Twitter, examining the relationships and interactions among users with differing political views.

### 2.3. Frameworks for Analyzing Social Media Data

Recent advancements in social media research have increasingly focused on integrating both user-generated content and user interactions to capture the complex dynamics of online communities. Dargin et al. introduced a framework that examines social media use during disasters, highlighting the role of socioeconomic factors and geographic location in influencing social media usage patterns during hurricanes (Dargin, Fan, and Mostafavi 2021). Their framework reveals significant digital divides that affect information access and reliability perceptions, emphasizing the need for equitable resilience in disaster response efforts. Similarly, Fan et al. developed a framework to investigate the diffusion patterns of situational information on social media, distinguishing between the roles of regular users (crowd) and influential users (hubs) in spreading information (Fan et al. 2020). Their study on Hurricane Harvey demonstrated that early intervention by hub users significantly boosts the speed and magnitude of information dissemination, providing practical strategies for improving emergency response through social media.

In the realm of political content, frameworks such as TwiFly, presented by Antelmi et al. offer real-time monitoring and analysis of political discourse on platforms like Twitter. TwiFly captures and visualizes data related to followers, tweets, retweets, and mentions, providing insights into political strategies and public sentiment during elections (Antelmi et al. 2018). This framework has been applied to analyze political trends and election results, demonstrating its utility in political analysis and forecasting. Babvey et al. emphasized the role of social media in political polarization and discourse fragmentation, highlighting the need for comprehensive frameworks that can capture the nuances of online political interactions and their implications for public opinion (Babvey et al. 2020). Their work underscores the importance of understanding how social media shapes political discourse and influences voter behavior.

Frameworks for disaster response have also been developed to utilize social media for real-time situational awareness and infrastructure resilience. Fan et al. proposed a systematic framework for detecting infrastructure-related topics during disasters using social sensing and text mining. Their framework integrates keyword search, text lemmatization, POS tagging, TF-IDF vectorization, LDA topic modeling, and K-means clustering to analyze tweets related to infrastructure performance. Applied to Hurricane Harvey, the framework effectively summarized and tracked infrastructure-related topics, providing actionable insights for improving infrastructure resilience and disaster response strategies (Fan et al. 2018). Rajput et al. modeled inter-organizational communication networks on social media during disasters, using Hurricane Harvey as a case study (Rajput et al. 2020). They identified the roles of different organizations in disseminating situational information and highlighted the importance of government and non-government collaboration in disaster response efforts.

Moreover, Zhang et al. reviewed the use of social media for public information and warning during disasters, proposing an interdisciplinary approach that leverages the participatory nature of social media to enhance disaster management strategies (Zhang et al. 2019). Their framework identifies key functions such as acquiring situational awareness, supporting peer-to-peer help activities, and enabling disaster management agencies to hear from the public. This comprehensive approach underscores the potential of social media to improve communication and coordination during emergencies. Additionally, the work of Lipizzi et al. on integrating machine learning and social network analysis to understand social media dynamics further highlights the need for frameworks that can generalize across different domains and provide deeper insights into the impact of real-world circumstances on social media interactions (Lipizzi, Iandoli, and Marquez 2016).

According to (Arnold and Wade 2015), systems thinking is defined as a set of synergistic analytical skills used to enhance the ability to identify and comprehend systems, predict their behaviors, and devise modifications to achieve desired outcomes. This definition highlights the significance of understanding the interconnections within a system and the dynamic behaviors that result from these interactions. Numerous studies in the literature have employed this methodology and perspective to address various problems. For instance, (Kalantari et al. 2023) emphasizes the importance of modeling human interactions and information diffusion within driver collaboration networks. By employing a dynamic overlapping community detection (DOCD) algorithm, researchers can identify and track communities and their leaders over time, allowing for a more comprehensive understanding of network dynamics and information diffusion, leading to better decision-making and optimization strategies. In management systems, (Maleki, Yaghoubi, and Fander 2022) focuses on analyzing the supply chain not as isolated components but as a cohesive system. Their methodology involves developing mathematical models to simulate real-world scenarios and examining the effects of different variables (e.g., organic levels, sales efforts) on the supply chain's performance. In engineering systems, (Amini, Marsooli, and Neshat 2024) addresses the complex problem of calibrating vegetation drag coefficients in coastal environments. Similarly, (Neshat et al. 2024) adopts a holistic, interconnected, adaptive, and multi-objective approach to optimize hybrid wave-wind energy systems. Even in cutting-edge research utilizing quantum computing-inspired optimization, the same systems thinking approach is evident (Amani and Kargarian 2024; Amani, Mahroo, and Kargarian 2023). There is more research with this perspective (Khameneh et al. 2023; Perez-Pereda, Krstikj, and Ramirez-Marquez 2024), which shows that this perspective enables researchers to analyze the system in detail and understand the behavior of its constituent components while also taking a holistic view to comprehend the system's behavior over time within the context of larger systems. This approach has been instrumental in shaping our framework.

In summary, our literature review reveals that recent frameworks increasingly integrate the analysis of both user-generated content and interactions (networks of replies, retweets, users, etc.). These frameworks vary in focus, with some emphasizing specific domains like political tweets or disaster-related tweets, others examining user interactions based on shared or divergent ideas, and some analyzing conversation networks and interactions in response to controversial tweets. Additionally, certain studies employ predefined network measures to represent the dynamics of complex systems. However, there remains a clear need for a generalizable framework that leverages fundamental social science theories to present the dynamics of discourse interactions and fragments and to elucidate the impact of real-world circumstances on social media, as well as the influence of social interactions on societal behavior. Our research aims to fill these gaps by introducing a comprehensive framework focused on major discourses and discourse fragmentation analysis. This approach offers insights into the direction and intensity of the impact of real-life circumstances on society, reveals how people react to different discourse fragments, and uncovers patterns of growth, decline, cohesion, and polarization within these fragments.

This study aims to contribute to urban sociology and community studies by examining the relationship between online discourse, particularly hate speech, and urban social dynamics. By analyzing social media data from New York City, San Francisco, and Seattle during significant events in 2020, we explore how online communities interact with offline urban social structures. Our approach, combining text classification and social network analysis, offers a perspective on community formation in the digital age. By focusing on online hate speech in urban contexts, we hope to provide insights into how digital interactions may reflect and potentially influence urban environments. This research may have implications for understanding contemporary urban communities and could inform discussions on urban policy-making and community engagement in increasingly digitally-connected cities.

The following summarizes the key contributions of our research:
- **Developing an Integrated Framework:** We designed a novel framework that combines text classification with social network analysis to effectively identify and understand online communities.
- **Temporal Visualization of Community Dynamics:** We created a method for temporal visualization to capture and represent the evolution of online communities over time.
- **Introducing Key Analytical Elements:** Our research introduced fourteen critical elements derived from the structure and dynamics of online social media, providing comprehensive insights into social interactions and discourse patterns.

- **Facilitating Discourse Fragmentation Analysis:** We established a framework that enables decision-makers to identify primary discourses and their fragments, enhancing the analysis of discourse fragmentation and its social implications.
- **Conducting a Real-life Case Study:** We applied our framework to a case study using real-world data, demonstrating its effectiveness in analyzing the formation and development of online communities.

Our work contributes significantly to the literature by providing a robust tool for analyzing the complex dynamics of online discourses, thereby offering valuable insights into the interplay between social media interactions and societal behaviors.

## 3. Methodological Framework

This section details the methodological framework employed to investigate the formation and evolution of communities within social networks. The framework, illustrated in Figure 1, consists of four interconnected components, each encompassing distinct processes that will be elaborated upon in the subsequent subsections.

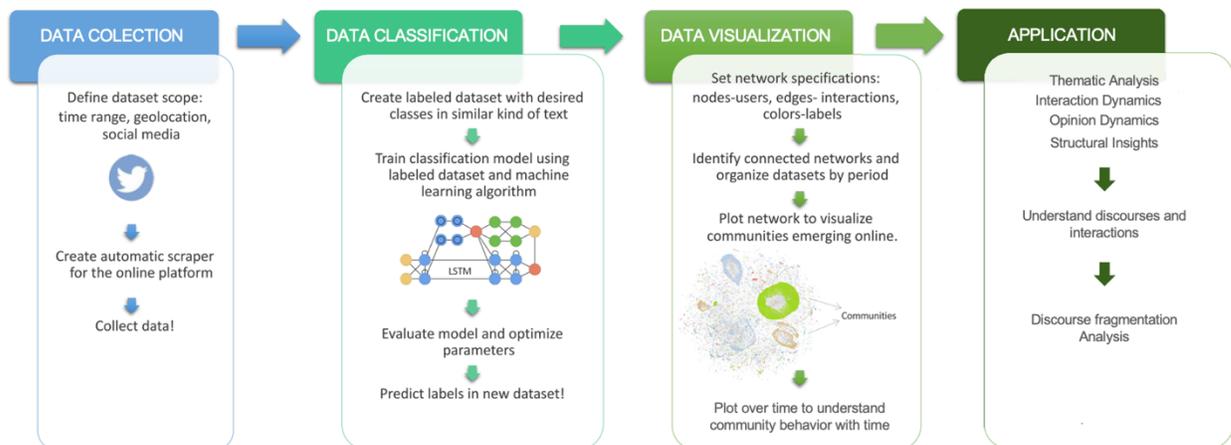

Figure 1. Conceptual framework developed for understanding community formation

### 3.1. Data Collection

Robust data collection forms the cornerstone of data-driven research. In this study, the focus is on understanding the temporal evolution of online communities, which necessitates data that captures change over time. Therefore, the collected dataset must incorporate timestamps indicating the precise moment of data creation.

A community, within this context, is defined as an aggregate of individual social media users and their collective actions. Departing from prior studies that emphasized group behavior, this research aims to identify individual users within their specific geographic contexts. Consequently, location information is a critical component of the dataset. This focus allows for the investigation of how online communities form and develop around significant circumstances, with cities serving as the primary lens for observation.

Following the definition of the dataset's key attributes and the establishment of the study's scope, an automated procedure for data collection is implemented. This process is developed to gather Twitter tweets based on the following criteria:
- Starting with a list of geographical areas $A$, each area $a_n$ is defined by querying potential neighborhoods and aliases within that area. For instance, if $a_1$ = "Los Angeles", the corresponding query for $a_1$ would be: ("Los Angeles") AND (("Los Angeles County") OR ("Orange County") OR ("Riverside County") OR ("Ventura County") OR ("Long Beach") OR etc.).
- When an circumstance $s^t$ takes place on date $t$, the data collection period spans from $t - \Delta_a$ to $t + \Delta_b$, where $\Delta$ ranges between 1 and 30 days.
- Additional filters can be applied to minimize irrelevant data; for example, if all geographical areas in $A$ are in the United States, the query would include: AND (country: "US") AND (language: "en").
- Each data entry necessitates the inclusion of essential attributes such as "created_at" to denote the precise time of tweet publication in Unix format, "user_id" to identify the author of the tweet, the "tweet" itself to capture the textual content of the post, and a unique "id" to serve as a distinctive identifier.

Due to the sheer volume of daily social media activity, capturing all posts for a given city within a defined timeframe is infeasible, particularly for densely populated areas. Sample size selection must consider the level of engagement within the city during the circumstance under investigation. Previous studies have often utilized samples of approximately 20,000 posts per week, which aligns with the download limitations imposed by many popular social media APIs.

The importance of sample size is amplified in the "Data Visualization" phase of this research, as it directly impacts the construction of the social network, which is dependent on user interactions. If the sample size is too small, resulting in insufficient interactions, it may not adequately represent the dynamic changes occurring within the community, necessitating further data collection iterations. A more in-depth discussion of network creation is presented in subsequent sections.

### 3.2. Topic Classification

Text classification, a cornerstone of Natural Language Processing (NLP), aims to categorize textual data into predefined classes or labels. This supervised learning process necessitates a carefully constructed training dataset that accurately represents the data to be classified. Ideally, this dataset mirrors the structure and content of the target data, with each observation meticulously labeled by human experts or through crowdsourced efforts (Chang, Amershi, and Kamar 2017). While advancements in machine learning have led to the availability of numerous labeled datasets, selecting appropriate training data is paramount. Utilizing unsuitable training sets can result in inconsistent, misleading, or biased outcomes.

Machine learning models are not without limitations, and researchers must carefully consider these constraints. The "Discussion" section will address the ethical implications associated with these limitations. In the present context, several crucial factors must be considered when selecting or constructing training datasets for NLP tasks:

- **Structural Alignment:** The textual structure of the training data should closely resemble the structure of the data to be classified.
- **Linguistic Consistency:** Maintaining a consistent language throughout the process is essential. While automatic translation tools may be necessary for less common languages, they can introduce errors, particularly when dealing with slang and ephemeral expressions.
- **Contextual Relevance:** The context in which the data was generated should be thoroughly understood and evaluated for its applicability to the current task.
- **Source Reliability:** When utilizing external training datasets, researchers must ensure the reliability of the source and have a clear understanding of the mechanisms employed for the labeling process.

While not exhaustive, these considerations form a foundation for data reliability. In the context of this framework, classes should represent distinct groups identifiable within an online community, such as political affiliations or news topics of interest.

Once a suitable training dataset is obtained, the next step is to select an optimal classification model. The literature highlights the superior performance of deep learning models over traditional machine learning methods in numerous text classification tasks (Luppescu and Romero 2017) However, even within the realm of deep learning, model performance can vary significantly depending on the specific task and the nature of the data. To determine the most suitable model, an empirical approach is adopted, involving evaluating several models and selecting the one with the best performance according to the metric that is used based on the task.

This study addresses a multi-class text classification problem, for which neural network models like Long Short-Term Memory (LSTM) (Hochreiter and Schmidhuber 1997; Luppescu and Romero 2017) and transformer-based models like BERT (Devlin et al. 2018; González-Carvajal and Garrido-Merchán 2020) have demonstrated remarkable effectiveness.

Before fitting the data to a chosen model, preprocessing is necessary. For both LSTM and BERT, this involves cleaning the data (removing links, punctuation, numbers, and stop words), lemmatizing the text, and creating a matrix of word embeddings. The labeled dataset is then partitioned into an 80% training set and a 20% testing set, with classes serving as target variables and word embeddings as features. Models are trained on the training set and then evaluated on the testing set to generate predictions. The accuracy of the model is assessed by comparing the predicted classes to the original classes. This process is iterative, with model parameters adjusted and the model retrained until optimal accuracy is achieved. The model with the highest accuracy is then selected to classify the unlabeled data, assigning a class to each social media post. The final output is the original dataset augmented with the predicted "class" target variable. The set of assigned classes in the dataset is denoted as $C = \{c_1, c_2, c_3, \ldots, c_k\}$.

### 3.3. Visualization methodology

To visually represent the dynamics of user interactions, a subset of the dataset, denoted as $W\{a, s^t\}$ is transformed into a graph structure. In this graphical representation, users are represented as nodes, and the interactions between them (replies, reposts, and mentions) are depicted as edges. Each post's assigned class remains associated with the user node that originated the post, and a distinct color is used to visually distinguish each class.

The graph $G$, representing a specific area during a particular circumstance, is formally defined as $G\{a, s^t\} = (U, N, C)$, where:

- $N$ denotes the set of interactions: $N = \{n_1, n_2, \ldots, n_p\}$, with $i = (u_a, u_b) \in W\{a, s^t\}$.
- $C$ signifies the set of classes for each "sender" node $u_a$ and the corresponding tweet class: $C = (u_i, c_j)$, where $u \in U$ and $c \in C$. Nodes acting as ``receivers" are not assigned classes. If a node appears as a "sender" in multiple interactions, it is counted as a single node if all interactions share the same class $c$, or as multiple interactions if $u$ has different classes $c$ in $W\{a, s^t\}$ on day $t$.

The expression of emotions and opinions on social media platforms, often referred to as "online venting," is a widespread phenomenon. However, for these individual expressions to significantly impact a community, they typically require engagement or interaction from other users. In this context, we posit that posts that fail to generate any interaction are unlikely to play a substantial role in shaping community dynamics.

To enhance the clarity and interpretability of the network visualization, we exclude nodes that lack labels, as these represent users who have not actively participated in the discourse. This filtering process reduces noise and focuses the analysis on those users who are actively contributing to the formation and evolution of online communities.

Figure 2 illustrates a social network graph where distinct communities are clustered and differentiated by color. While larger communities are readily apparent, smaller groups and even isolated individuals can be observed within the broader network structure. This is due to the fact that classification is based on post content rather than solely on interaction patterns. Notably, Figure 2 reveals the presence of a small blue group nested within a larger green group, illustrating the concept of a community (small hub) existing within a larger community (bigger hub).

Social media is often a platform for individuals to express emotions and opinions. To influence the community, posts should generate interactions. Thus, posts without interactions are filtered out to enhance visualization and reduce noise. Figure 2 illustrates a social network with clustered communities represented by different colors. Larger groups represent distinct communities, but individuals or small groups are also visible, classified by post content rather than group interaction. This is seen in the small blue group within the larger green group at the bottom right of Figure 2. The blue group (small hub) is a community within a community (larger hub).

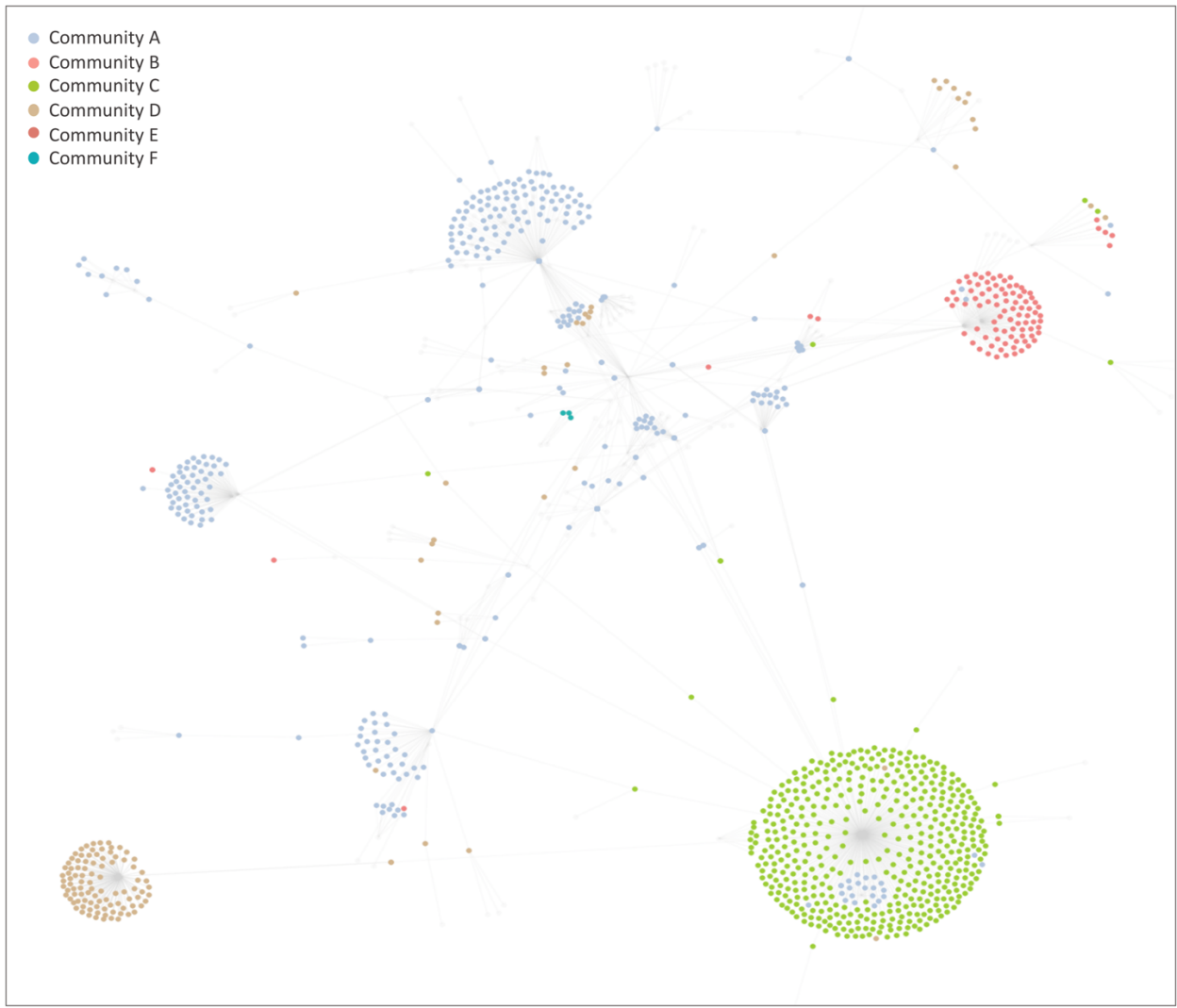

*Figure 2. Community-oriented Social Network*

The dataset subset $W\{a, s^t\}$ is drawn from the data gathered for each area $a$ over a specified time frame $t$, where $t$ falls within the range $(t_{s-d}, t_{s+d})$$(t_{s-d}, t_{s+d})$$, indicating the period $d$ days before and after the circumstance $t_s$. This method is applied consistently across all circumstances and areas, resulting in around $2d + 1$ static visual representations of these communities over time. By arranging these visualizations chronologically and juxtaposing them according to circumstance, area, and time, we can construct a comprehensive understanding of the dynamic formation and evolution of these communities within a broader context.

### 3.4. Discourse Fragmentation Analysis

Leveraging Twitter data offers profound insights into the dynamics and characteristics of social media. This analysis identifies fourteen key elements that can highlight broader societal trends and insights. To provide a structured and theoretically grounded understanding, these elements are categorized into four main groups: Thematic Analysis, Interaction Dynamics, Opinion Dynamics, and Structural Insights.

#### 3.4.1. Thematic Analysis

The thematic analysis involves understanding the content and themes of discussions, reflecting the central topics driving engagement and societal concern.

- Content Focus: Identifying the central themes or topics of discussions allows us to understand the primary drivers of engagement and attention on Twitter. This element focuses on what people are talking about and why these topics gain traction, serving as a mirror to contemporary societal interests.
- Societal Discourse Themes: Identifying overarching topics and themes driving public discourse provides insights into the focal points of societal concern and interest. This element helps to map out the larger conversations shaping public opinion and societal priorities.
- Cultural and Societal Trends: Examining how cultural shifts and societal trends manifest in social media discussions provides a reflection of changing societal values and norms. This analysis can reveal deeper cultural transformations and emerging societal narratives.

#### 3.4.2. Interaction Dynamics

Interaction dynamics focus on the structure of user interactions, the flow of information, and the relational aspects of social media networks.

- Relational Dynamics: Exploring the structure of user interactions reveals how users are connected and how information flows between them. This element is crucial for understanding the network properties of social media, including patterns of influence and communication.
- Influential Participants: Pinpointing key users who initiate and influence discussions helps understand their impact on shaping the narrative. This element focuses on the role of influential actors in driving and moderating social media conversations.

### 3.4.3. Opinion Dynamics

Opinion dynamics involve the formation, evolution, and polarization of opinions within the social media landscape.
- Opinion Formation and Evolution: Tracking how opinions on specific topics develop and change over time provides a dynamic view of shifting perspectives and trends. This element highlights the processes through which public opinions are formed and transformed.
- Polarization: Analyzing how divergent views become more extreme as like-minded individuals reinforce each other's beliefs shed light on the polarization of opinions. This element examines the social mechanisms that lead to increased ideological divides.
- Segmentation: Observing the formation of different groups around specific topics or beliefs reveals how limited exposure to alternative views creates segmented communities. This element focuses on the fragmentation of public discourse and the creation of echo chambers.
- Ephemerality: Noting the short lifespan of topics, which are quickly replaced by the next trending issue without deeper engagement or resolution, highlights the transient nature of social media discussions. This element emphasizes the rapid and often superficial nature of online discourse.
- Historical Comparisons: Comparing current social media trends with past data helps identify historical changes in public opinion and societal priorities. This temporal analysis offers a dynamic view of how social concerns and discussions evolve over time.

### 3.4.4. Structural Insights

Structural insights focus on the overarching characteristics of social media discussions and their broader implications.
- Reaction to Social and Political Issues: Exploring how users react to significant circumstances, changes, or societal movements provides a snapshot of public sentiment. This element captures the immediate and collective response of the social media populace to external stimuli.
- Dominance: Recognizing how certain voices or topics overpower others, often due to algorithmic amplification or sheer volume, highlights the asymmetries in social media influence. This element examines the structural inequalities in the visibility and impact of different participants and topics.

- Diversity and Echo Chambers: Assessing the extent of cross-cutting dialogues that bridge different groups versus conversations confined predominantly within closed networks reveals the degree of ideological diversity and insularity. This element explores the balance between exposure to diverse perspectives and the reinforcement of existing beliefs.
- Discussion Cohesiveness: Identifying topics that foster cohesive versus fragmented discussions offers insights into the stability and uniformity of conversations. This element assesses the extent to which discussions are integrated or divided, reflecting the coherence of public dialogue.

These fourteen elements are meticulously crafted to offer a comprehensive perspective on how social media data can reflect broader societal changes and function as a barometer for public sentiment across a wide spectrum of topics. By categorizing these elements into thematic, interactional dynamics, opinion-based dynamics, and structural insights, we present a theoretically robust framework for comprehending the intricate dynamics of social media. Upon deeper examination of the core principles of our proposed visualization methodology, it becomes evident that our approach offers a better tool for elucidating the opinion-based and structural insights segments of our key elements. Therefore, we prioritize and emphasize these components in our analysis.

Discourse fragmentation is a term that captures the splintering of public conversation into smaller, often isolated sub-discussions or discourses. It is characterized by the lack of a central narrative or shared understanding, leading to a scenario where multiple conversations occur in parallel, with limited interaction or integration between them. In modern society, especially online, this is often amplified by algorithmic curation, selective exposure to media, and the echo chamber effect. Focusing on our problem in social media, we can see that discourse fragmentation is understandable through opinion dynamics and structural insights, as defined previously. Therefore, our final effort is to present our application and analysis layer as a discourse fragmentation analysis tool, which can be invaluable for major players in politics, social sciences, and governance.

The idea of introducing several key elements that can highlight broader societal trends and insights from raw data and then using these elements to obtain a higher level of understanding of the theme of our complex system, and in its interaction dynamic, opinion dynamic, and structure, and finally using these higher levels of understanding to analyze the discourse fragmentation and its impact on community shaping roots from systems thinking principles and best practices which we discussed in literature review. Figure 3 shows the underlying idea for our analysis.

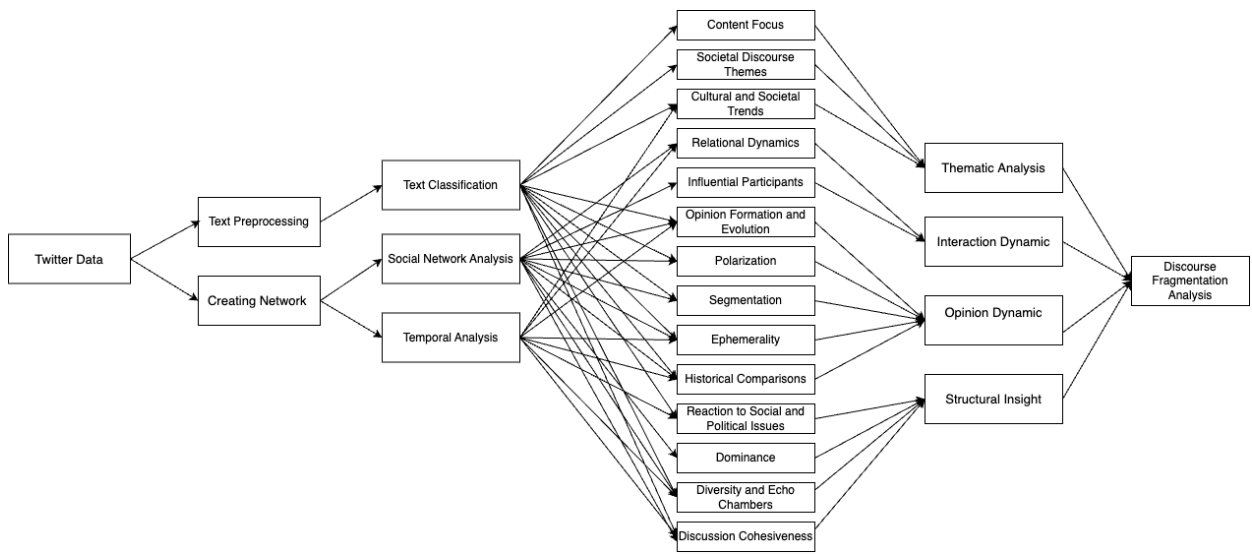

*Figure 3. Workflow of Discourse Fragmentation Analysis for Twitter Data*

## 4. Experiment: Online Hate Groups

This paper demonstrates the utilization of the suggested methodology to examine dynamic social networks with a community focus. Babvey et al. explored the consequences of lockdown measures on children's encounters with harmful content, emphasizing the significant influence online platforms can have on people's well-being (Babvey et al. 2020). However, their study also underscored the intangible nature of online communities, particularly for authorities trying to pinpoint individuals responsible for spreading harmful online behavior like the dissemination of false information, online harassment, prejudiced language, or the encouragement of violent acts. From an ethical standpoint, classifying a user as a possible perpetrator of online offenses solely based on a single social media tweet, especially when utilizing artificial intelligence, raises significant concerns. Hence, this research refrains from singling out specific individuals, and usernames will be kept private, and the focus is on illustrating the temporal behavior of users engaging in various forms of online hate speech within the context of real-life circumstances.

### 4.1. Data Acquisition

Twitter, a widely used platform for textual data and social network studies, was chosen due to its readily available metadata, encompassing user identifiers, location information, timestamps, URLs, images, and references to other users. For this investigation, tweets originating from New York City, Seattle, and San Francisco were gathered in real-time over the course of 2020 and subsequently filtered by date in proximity to three significant national circumstances:

- The official designation of Coronavirus as a global pandemic on March 11, 2020, leading to widespread shutdowns and restrictions on mobility. The data collection period for this circumstance spanned from March 2nd to March 29th, 2020.
- The tragic killing of George Floyd by a police officer on May 25, 2020, which ignited the "Black Lives Matter" movement protesting police brutality. The data collection period for this circumstance spanned from May 18th to June 14th, 2020.
- The highly contested presidential election on November 3, 2020, occurring after a four-year term marked by Donald Trump's presidency and widespread critique of his handling of the pandemic. The data collection period for this circumstance spanned from October 19th to November 14th, 2020.

### 4.2. Topic Classification

To identify groups disseminating harmful language online, tweets need to be classified based on the nature of the language employed. Hate speech encompasses a variety of intentions and expressions, so this research categorizes tweets by the type of harmful language used, thereby recognizing communities as groups engaging in such behavior. The annotated dataset utilized for training the model capable of assigning multiple classes was derived from prior research (Mollas et al. 2020), which utilized online comments and integrated machine learning techniques, human labeling, and validation through crowdsourcing. The eight categories of harmful language used are:
- Sexism for harmful language targeting gender,
- Racism for language expressing racial bias,
- Xenophobia for language expressing prejudice against individuals based on their country of origin,
- Ableism for language discriminating against individuals with disabilities,
- Homophobia for harmful language directed at individuals based on their sexual orientation,
- Religious Intolerance for harmful language targeting religious beliefs.

Two distinct models underwent training and assessment using this dataset ($W$). The BERT model attained an accuracy of 75%, whereas the LSTM model reached 86%. Given its superior performance, an LSTM model capable of handling multiple classes was then utilized to assign each tweet within the collection a category from the predefined set of eight. Each category was assigned a score ranging from 0 to 1, calculated as a normalized weighted average of the annotators' votes. A tweet was classified as containing harmful language if at least one class received a score of 0.5 or higher. In cases where multiple classes met this threshold, the class with the highest score was selected. Following the classification process, tweets lacking any assigned classes were excluded, resulting in a final collection comprising 3,731,534 tweets. Figure 4 visually represents the distribution of tweets, user counts, and users exhibiting multiple forms of harmful language across the different cities.

| City | Tweets | Users | Multi-hate users |
|---|---|---|---|
| New York City | 1,139,236 | 457,503 | 105,904 |
| San Francisco | 1,305,772 | 567,557 | 131,700 |
| Seattle | 1,286,526 | 530,365 | 128,086 |

*Figure 4. Users engaging in hate speech tweets*

### 4.3. Visualization

The original dataset, $W$, is partitioned into nine distinct subsets denoted as $W\{a, s^t\}$, where $a$ represents the metropolitan areas (New York City, San Francisco, Seattle) and $s^t$ signifies the circumstances (Coronavirus pandemic, Black Lives Matter, US election). For each of these subsets, a corresponding graph $G\{a, s^t\}$ is generated. To enhance visual clarity and reduce noise caused by an abundance of nodes and edges, any nodes disconnected from the primary graph structure were removed, retaining only the most extensive connected subgraph, referred to as $G\{a, s^t\}$. Within the timeframe of each circumstance, the graph for a specific day exclusively incorporates tweets from that particular day. Figure 5 and Figure 6 depict the outcomes for New York City and Seattle during the Coronavirus pandemic, whereas Figure 7 and Figure 8 showcase the results for San Francisco during the Black Lives Matter movement and the 2020 presidential election, respectively. Supplementary information regarding community behavior in San Francisco at the onset of the pandemic, and in New York City and Seattle during the Black Lives Matter movement and the presidential election, can be found in the appendix.

These sequences illustrate the emergence of communities propagating various forms of harmful language at a frequent pace. Nevertheless, the transient nature of Twitter interactions often leads to these groups lasting for only a brief period, typically two to three days. For instance, in Figure 8, the initial quadrant (10/19) displays a blue community associated with xenophobic speech. On the subsequent day, this group diminishes in prominence as a new, comparable group takes shape. By the third day, the original group has dissipated, while the second group gains strength. On the fourth day, both groups have vanished entirely.

Typically, hate groups stand out distinctly against a neutral backdrop. However, in the context of the Black Lives Matter movement in San Francisco (Figure 7), the numerous red dots scattered across the graph, particularly during the initial period, signify the pervasive presence of racist language throughout the platform, not limited to specific communities. A similar trend is observed in Seattle during the Coronavirus pandemic (Figure 6), where xenophobic posts are widely distributed among users.

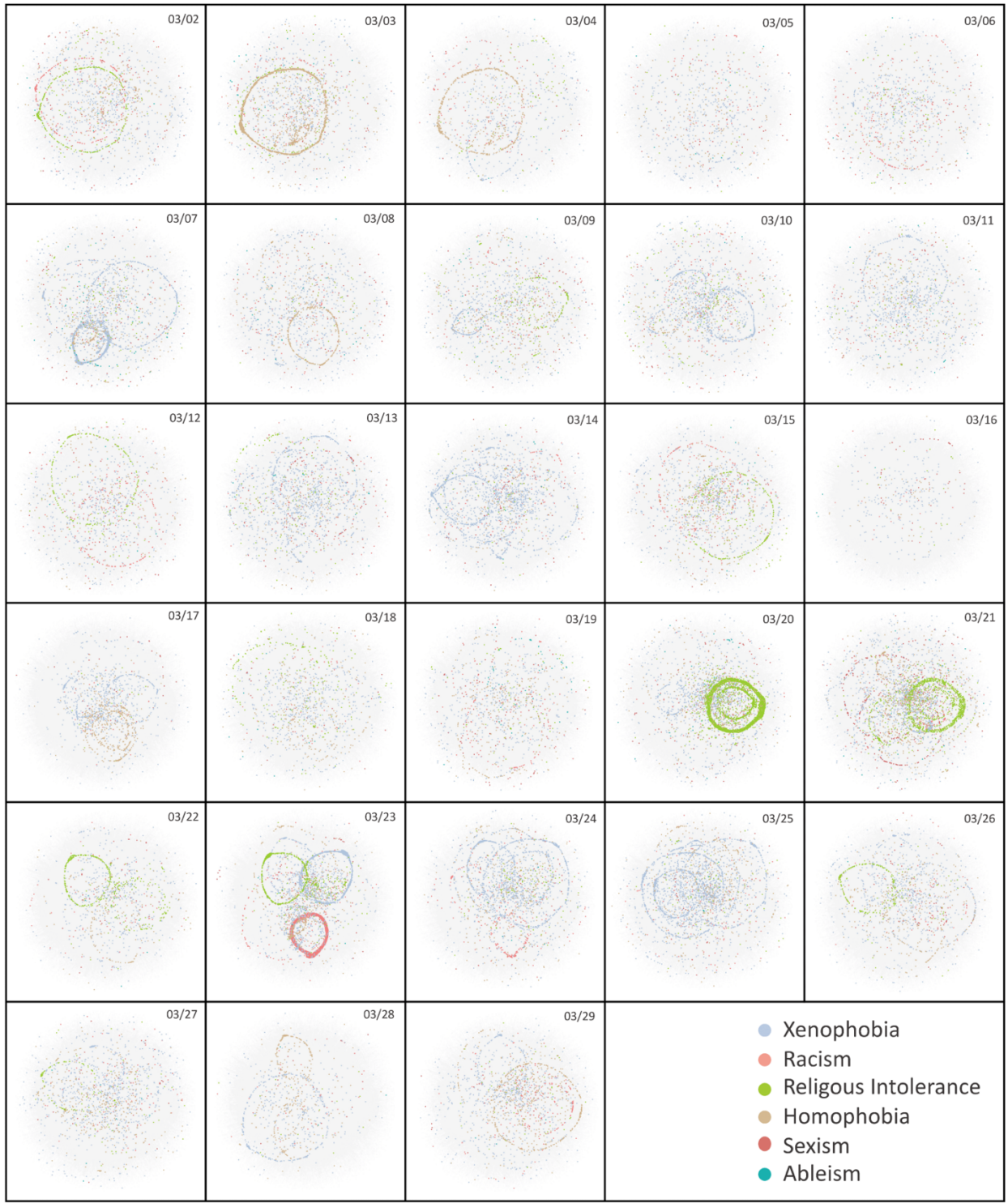

Figure 5. New York City during the Coronavirus pandemic

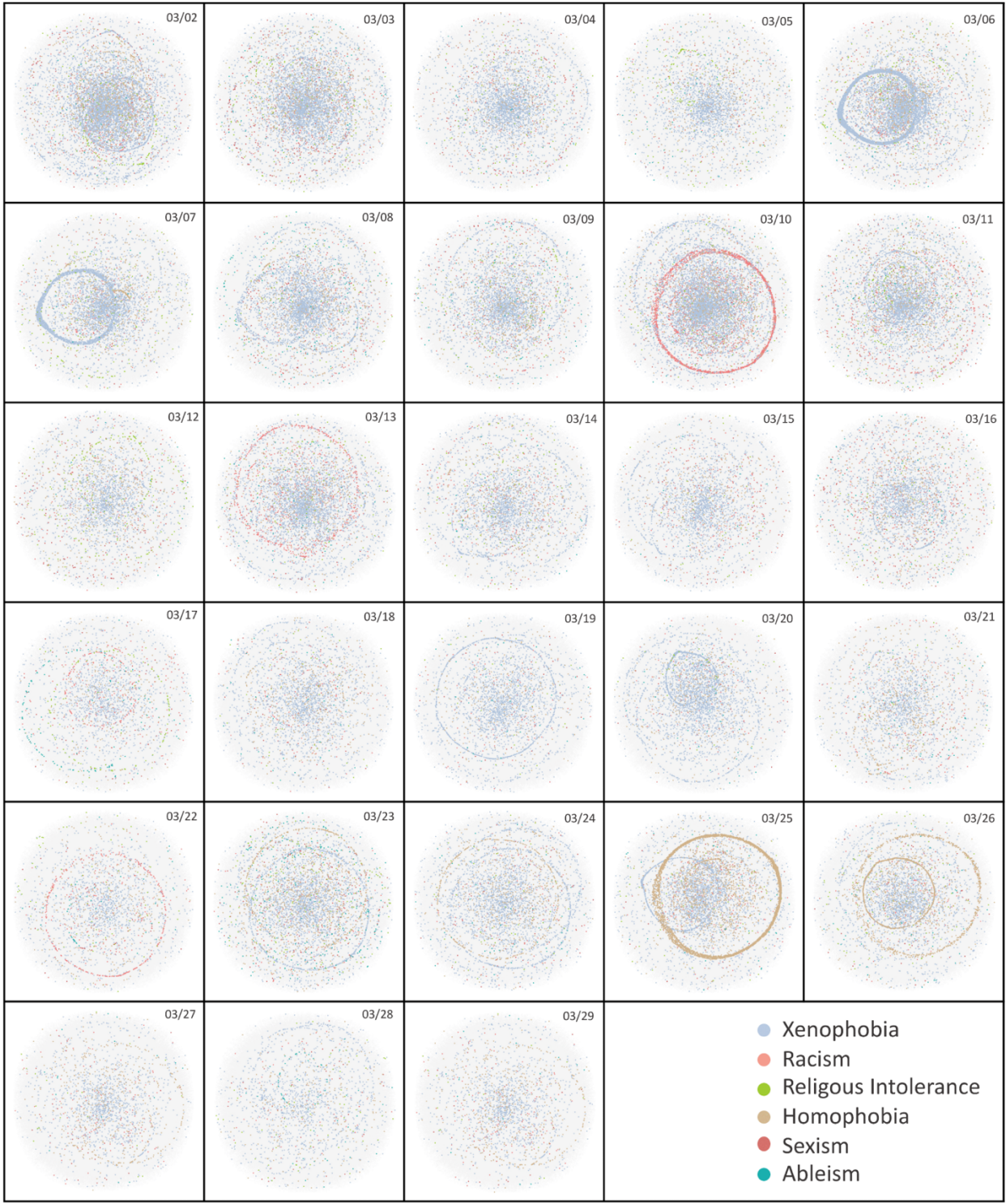

Figure 6. Seattle during the Coronavirus pandemic

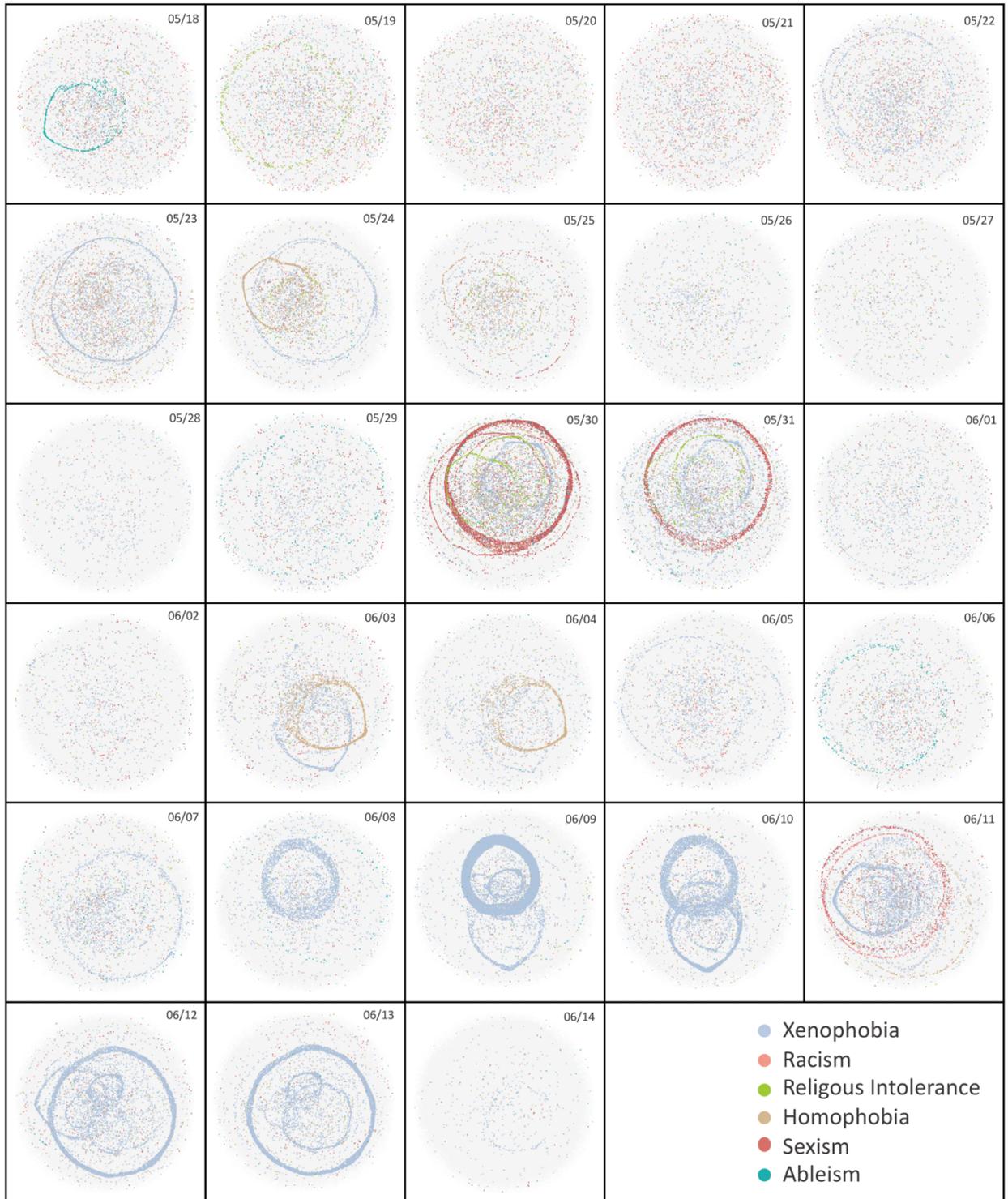

Figure 7. San Francisco in the Black Lives Matter movement

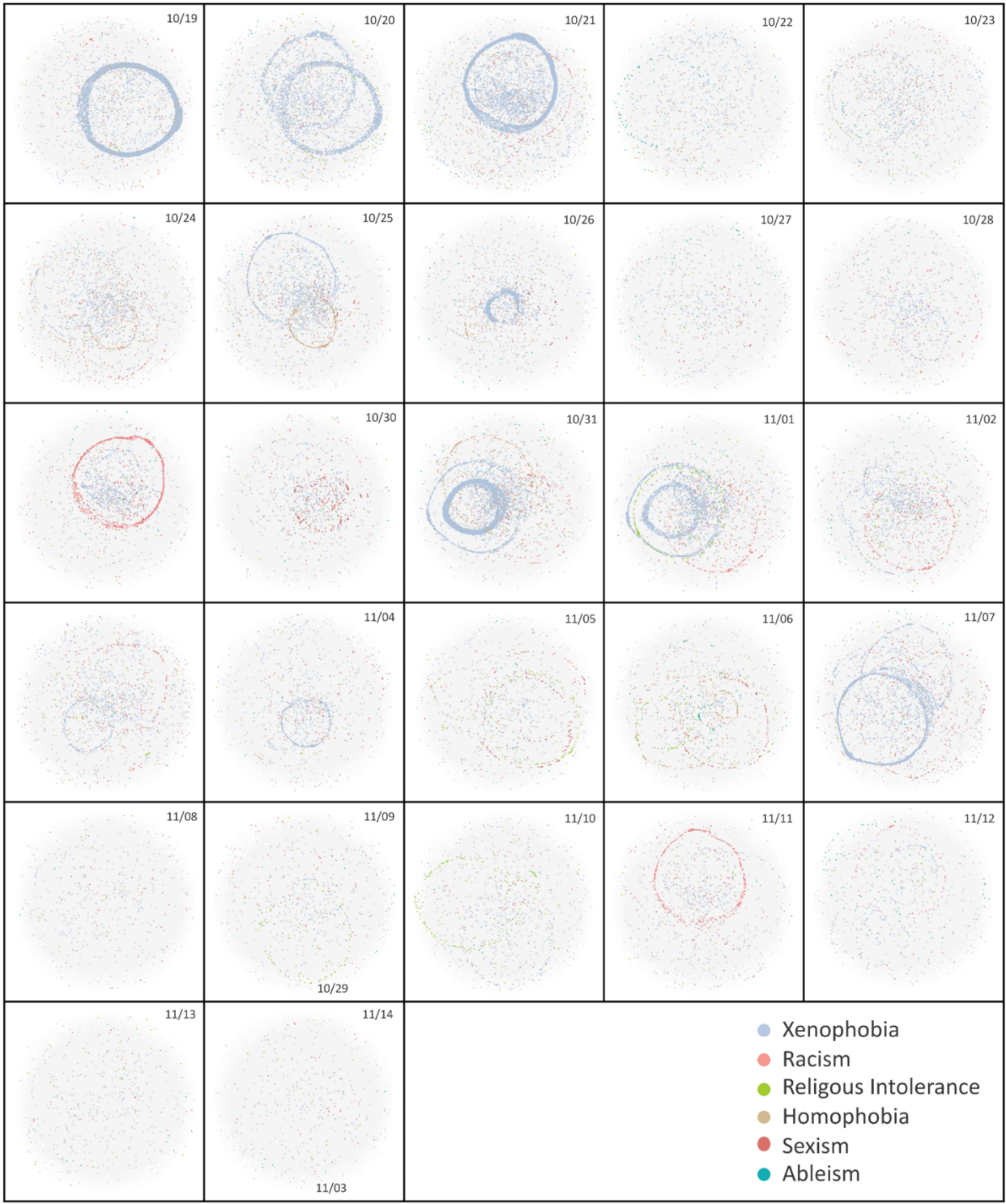

Figure 8. San Francisco in the 2020 presidential race

### 4.4. Discourse Fragmentation Analysis

The goal of the application and analysis layer of our framework is to help decision-makers gain a deeper understanding of how a primary discourse, such as discrimination and hate speech, and its fragments (sexism, racism, xenophobia, ableism, homophobia, and religious intolerance) evolve over time. It also examines the impacts of real-world circumstances on discourse fragmentation and the respective effects of different discourse fragments on the social system as a complex, living system. To achieve this goal, we introduced fourteen key elements that assist in thematic analysis, interaction dynamics analysis, opinion dynamics analysis, and structural insights. The following is a concise assessment of the capabilities linked to the application and analysis layer of our framework:

- **Opinion Formation and Evolution:** The proposed visualization methodology effectively captures the dynamic shifts in themes and trends. When utilized in real-time snapshots, it vividly illustrates the transitions and developments in opinion formation and evolving trends.
- **Polarization:** This visualization methodology reveals conflicting discourses and their interaction patterns, especially within inherently polarized discussions such as political debates. It elucidates the social mechanisms that contribute to increasing ideological divides.
- **Segmentation:** The methodology highlights the emergence of distinct groups centered around specific discourses, emphasizing the fragmentation of public dialogue and the formation of echo chambers.
- **Ephemerality:** By providing comprehensive insights into the historical evolution of public opinion and societal priorities, the methodology showcases the duration and lifespan of dominant discourses and the rapid emergence of new trending issues.
- **Historical Comparisons:** This methodology offers detailed insights into historical shifts in public opinion and societal priorities by comparing current and past social media trends. Analyzing multiple snapshots taken at different intervals highlights changes and developments over time.
- **Reaction to Social and Political Issues:** The methodology provides valuable insights into user responses to significant societal changes or movements, acting as a barometer that captures public sentiment over periods ranging from a few days to several months.
- **Dominance:** It illustrates how specific voices or topics overshadow others, often due to algorithmic amplification or significant volume. This highlights the

- imbalances in social media influence by examining structural inequalities in the visibility and impact of various participants and discussions.
- **Diversity and Echo Chambers:** The methodology adeptly highlights the scope of cross-cutting dialogues that bridge different groups, tracing connections between diverse social or political factions.
- **Discussion Cohesiveness:** When supplemented with coherency measures from NLP tools, this methodology effectively discerns topics that lead to cohesive or fragmented discussions. It provides valuable insights into the stability and uniformity of conversations.

From the analyses presented, we gain a deeper understanding of the reasons behind discourse fragmentation and its intricate patterns; analyzing discussion cohesiveness reveals the mechanisms through which specific discourses foster tighter engagement among participants. Consequently, echo chambers become evident when successive snapshots of the system indicate growth in particular discourses, as illustrated in figure 5 (Seattle during the Black Lives Matter movements), where echo chambers from June 8th contribute significantly to tweets centered around xenophobia. The dominance of certain discourses, as clearly depicted in the plots, underscores the power of each fragmented discourse and how they can absorb attention. This necessitates the contextualization of real-world circumstances with visual data to interpret public reactions, exemplified by the responses to the Black Lives Matter movement. Moreover, most fragmented discourses exhibit ephemerality, typically persisting for less than a day unless directly correlated with a real-world circumstance, an insight that our framework adeptly analyzes to elucidate the implications of discourse fragmentation. The segmentation of discussions by discourse emerges as a crucial strategy for dissecting discourse fragmentation, playing an indispensable role in shaping and evolving public opinion. Additionally, a focused analysis of political discourses enables a comprehensive examination of polarization, thereby yielding valuable insights into the broader phenomenon of discourse fragmentation. To further quantify discourse fragmentation, we calculated the number of distinct communities identified in each timeframe and assessed the degree of overlap between them. Our analysis revealed a significant increase in the number of distinct communities during periods of heightened social tension, suggesting a fragmentation of the discourse landscape. This fragmentation may contribute to the creation of echo chambers where harmful narratives are amplified and reinforced, hindering productive dialogue and exacerbating societal polarization.

5. Discussion

Figure 9 illustrates the distribution of classes across each class in the datasets. The three urban areas display similar percentages of tweets, potentially reflecting online behavior in the United States during 2020. It's worth noting that "Xenophobic" tweets were very common, making up almost 60% of all hate speech found in New York City. The second most common type was "homophobia." The second most prevalent category was "Homophobia".

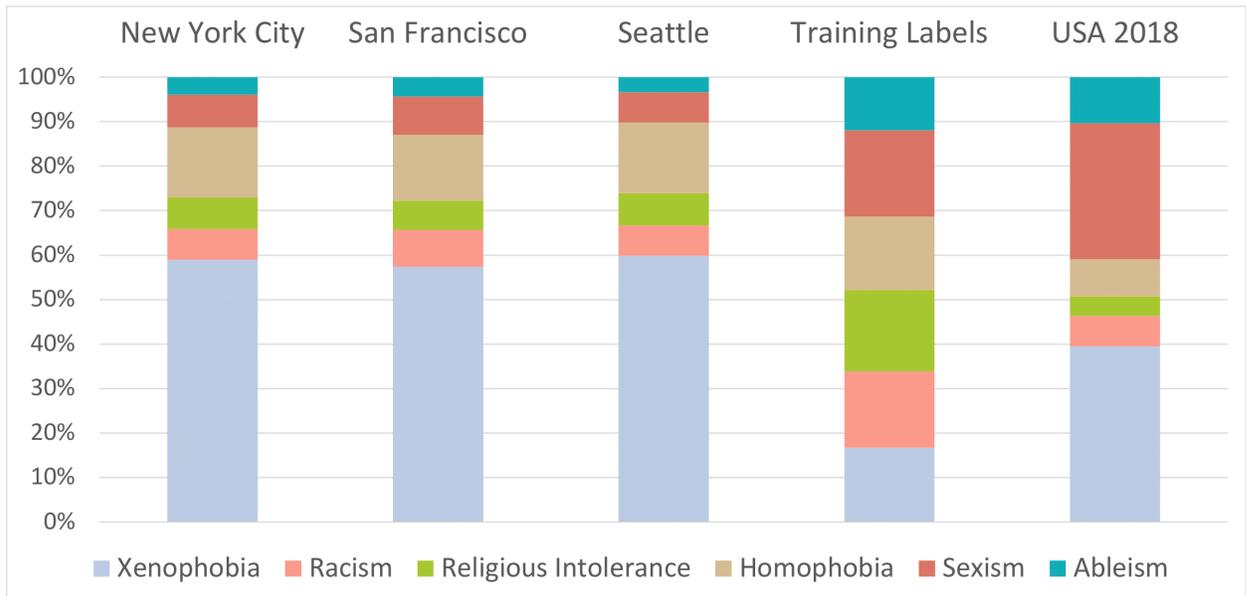

Figure 9. Proportion of classes for each hate speech category

However, these results differ significantly from a collection of tweets gathered from identical areas in 2018. In 2018, less than 40% of posts contained "Xenophobia", and the second most common category was "Sexism". This variation between the two years may be attributed to the distinct socio-political contexts. In 2018, the global #MeToo movement against sexual harassment gained momentum, leading to increased online discussions and backlash against the movement. Conversely, 2020 was marked by the devastating impact of the Coronavirus pandemic, during which the use of terms like "China virus" by prominent figures fueled xenophobia and anti-Asian sentiment. While overall crime rates have declined, a report from the United States Commission on Civil Rights (USCCR) documented a significant rise in anti-Asian hate crimes from 2019 through 2021 (United States Commission on Civil Rights 2023). This highlights the need for contextual analysis when interpreting crime statistics, as certain communities may experience disproportionate increases in targeted offenses despite broader trends.

Text classification models often grapple with mislabeling, a challenge exacerbated by the simplified assumptions of traditional machine learning approaches (Abdidizaji et al. 2024). While these models excel at optimizing metrics like precision or recall, they often struggle to capture the nuances of complex sociotechnical systems, particularly when dealing with concepts like fairness and discrimination. This limitation, as highlighted by Barocas and Selbst (Barocas and Selbst 2016), stems from an oversimplified understanding of the relationship between social context and technology. To address this, a more comprehensive approach is needed, one that considers the intricate interplay between social factors and technological systems.

Further, research by Suresh and Guttag (Suresh and Guttag 2019) reveals that many feature importance extraction methods in machine learning inadvertently perpetuate biases against marginalized groups. This is due to an implicit reliance on a singular, dominant perspective that overlooks the pluralistic, contextual, and interactional viewpoints essential to understanding fairness and discrimination. Empirical evidence from studies by Buolamwini and Gebru (Buolamwini and Gebru 2018) supports this claim, demonstrating that machine learning models frequently rely on features that may be spuriously correlated with classes during training.

Feminist epistemology critiques machine learning's feature importance extraction techniques, arguing they often embed biases that contradict the nuanced, situated, and interconnected perspectives valued by marginalized groups (Hancox-Li and Kumar 2021). This resonates with the broader critique within feminist technoscience, which emphasizes the ways in which technology can reflect and reinforce existing social inequalities (Burrell 2016). Furthermore, empirical investigations reveal a propensity for machine learning models to seize upon features that might exhibit only a spurious correlation with target classes during the training process (Buolamwini and Gebru 2018). A spurious correlation, in this context, refers to a statistical association between a feature and a class that is not based on a causal relationship, but rather on incidental co-occurrences in the training data. For instance, a model might associate the frequent use of the word "Immigrant" with negative sentiment due to its prevalence in negative contexts within the training data, even though the word itself is neutral. These spurious correlations can contribute to questionable classifications, as observed in some tweets classified as sexist or religiously intolerant in this study.

In conclusion, our study provides a novel framework for analyzing the dynamics of online communities and discourse fragmentation. By integrating text classification and social network analysis, we were able to identify and track the evolution of hate speech communities on Twitter. Our findings highlight the ephemeral nature of these communities, the impact of real-world events on their formation and dissolution, and the potential for discourse fragmentation to contribute to societal polarization. These insights have important implications for understanding the role of social media in shaping public discourse and for developing strategies to promote more inclusive and constructive online interactions.

## 6. Conclusion

The digital age has redefined the concept of community, transcending geographical boundaries and enabling individuals with shared interests to connect and form groups online. These virtual communities wield significant influence, shaping individual opinions and catalyzing real-world actions, both positive and negative. Understanding the dynamics of these online communities is imperative for policymakers, who must navigate the delicate balance between protecting freedom of expression and safeguarding individuals and society from potential harms.

In this paper, we present several key contributions:
- Designing a framework combining text classification and social network analysis to identify online communities.
-  Creating a temporal visualization to represent the changes in communities in the online environment over time.
-  Introducing fourteen key elements based on online social media structure and dynamics that encapsulate the insights expected from social structure and dynamics and analyzing the text and network combination results based on these elements.
-  Establishing a framework that allows decision-makers to identify main discourses and their fragments, thereby facilitating the analysis of discourse fragmentation and understanding its social implications in real life.
-  Developing a case study with real-life data to analyze the shaping of online communities.

Our case study focused on Twitter data collected in 2020 from three U.S. metropolitan areas. We categorized posts into classes such as "Sexist", "Racist", "Xenophobic", "Ableist", "Homophobic", and "Religiously intolerant". The analysis revealed that the lifespan of these identified online hate groups is typically short, ranging from one to three days. While circumstances like the Coronavirus pandemic amplified anti-Asian sentiment, the Black Lives Matter movement did not result in a similar surge in racist hate groups.

It is essential to acknowledge that not all users classified within these hate groups are actively engaging in harmful behaviors. Classifying subjective topics like hate speech using machine learning algorithms remains a nascent field, and caution is warranted regarding potential biases that can permeate various stages of the research process. Even sophisticated machine learning models can misinterpret posts by focusing on literal word meanings rather than grasping the nuances of contextual semantics.

The observed fragmentation of discourse raises concerns about the role of social media in shaping broader societal discourse. The formation of isolated communities and echo chambers can limit exposure to diverse perspectives, leading to the reinforcement of existing biases and beliefs. This can ultimately contribute to increased polarization and social division. Our findings highlight the need for further research into the mechanisms underlying discourse fragmentation and the development of strategies to promote more inclusive and constructive online conversations.

By enabling decision-makers to see the main discourses and their fragments, our framework provides a valuable tool for analyzing discourse fragmentation. This, in turn, allows for a deeper understanding of the social implications and real-life behaviors of these discourse fragments. Policymakers and researchers can use these insights to monitor and address the spread of harmful narratives and support the maintenance of healthy, constructive online communities.

For future research, this study could be expanded to analyze the continuity of these communities. Investigating whether large groups of users repeatedly appear over time could reveal their true membership or leadership within hate groups. Additionally, exploring the textual content of the tweets shared within these networks could lead to the development of an influence metric, identifying the most influential users in each discourse fragment. Finally, the proposed methodology could be extended to other domains, such as organizational sciences.

7. **Appendix**

Additional visualizations from the experiment:

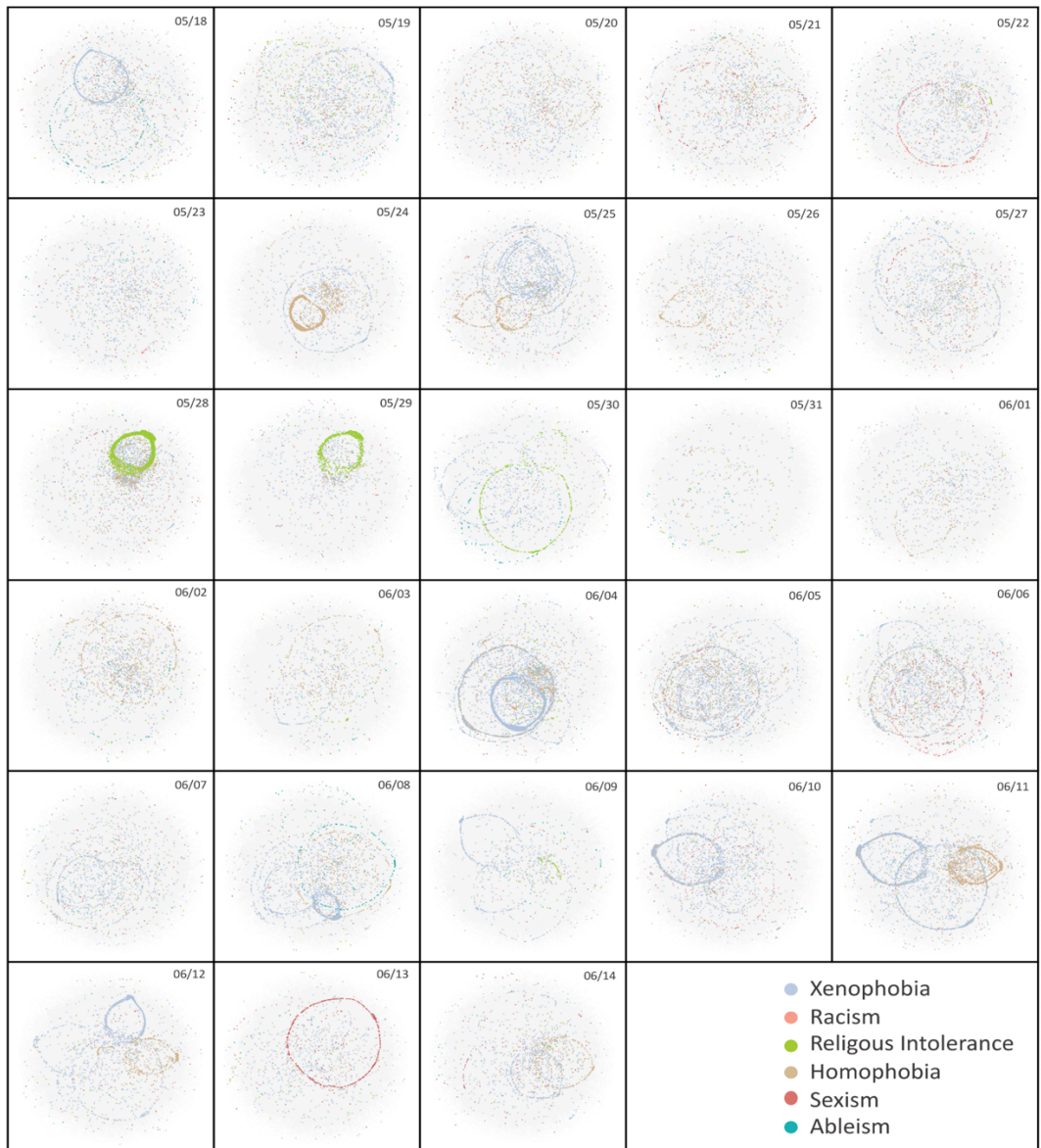

*Figure 10. New York City in the Black Lives Matter movement*

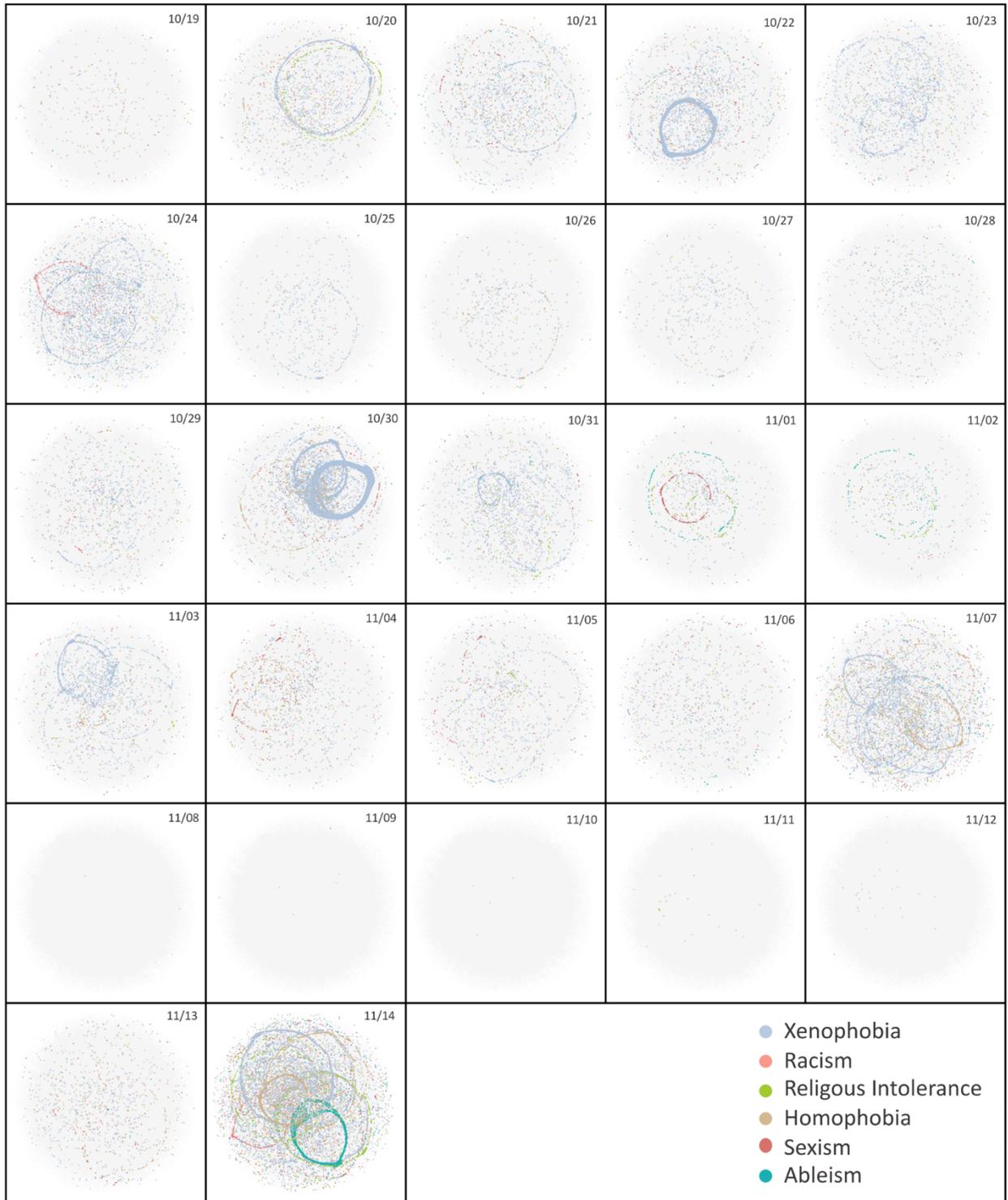

Figure 11. New York City in the 2020 presidential race

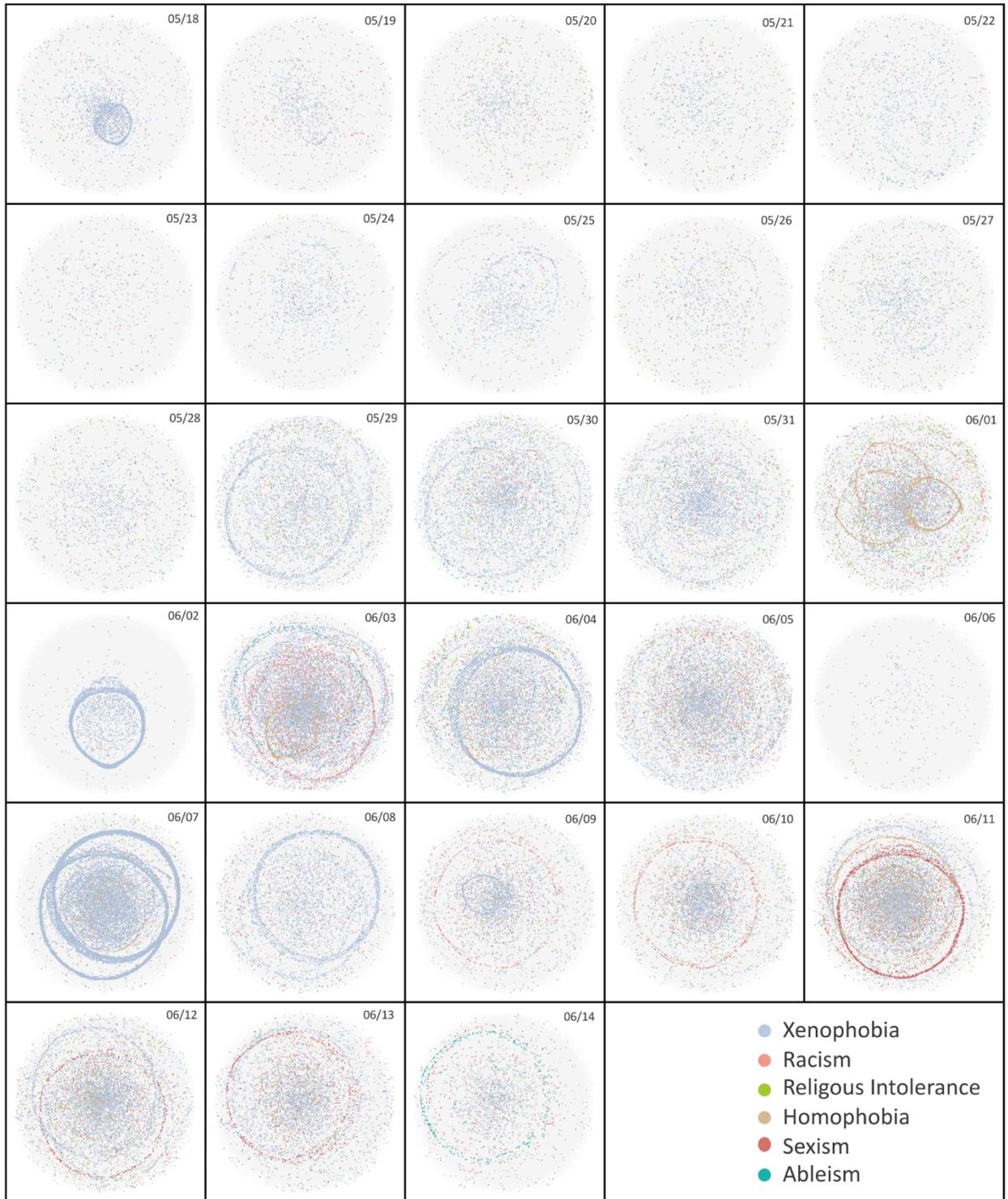

Figure 12. Seattle in the Black Lives Matter movement

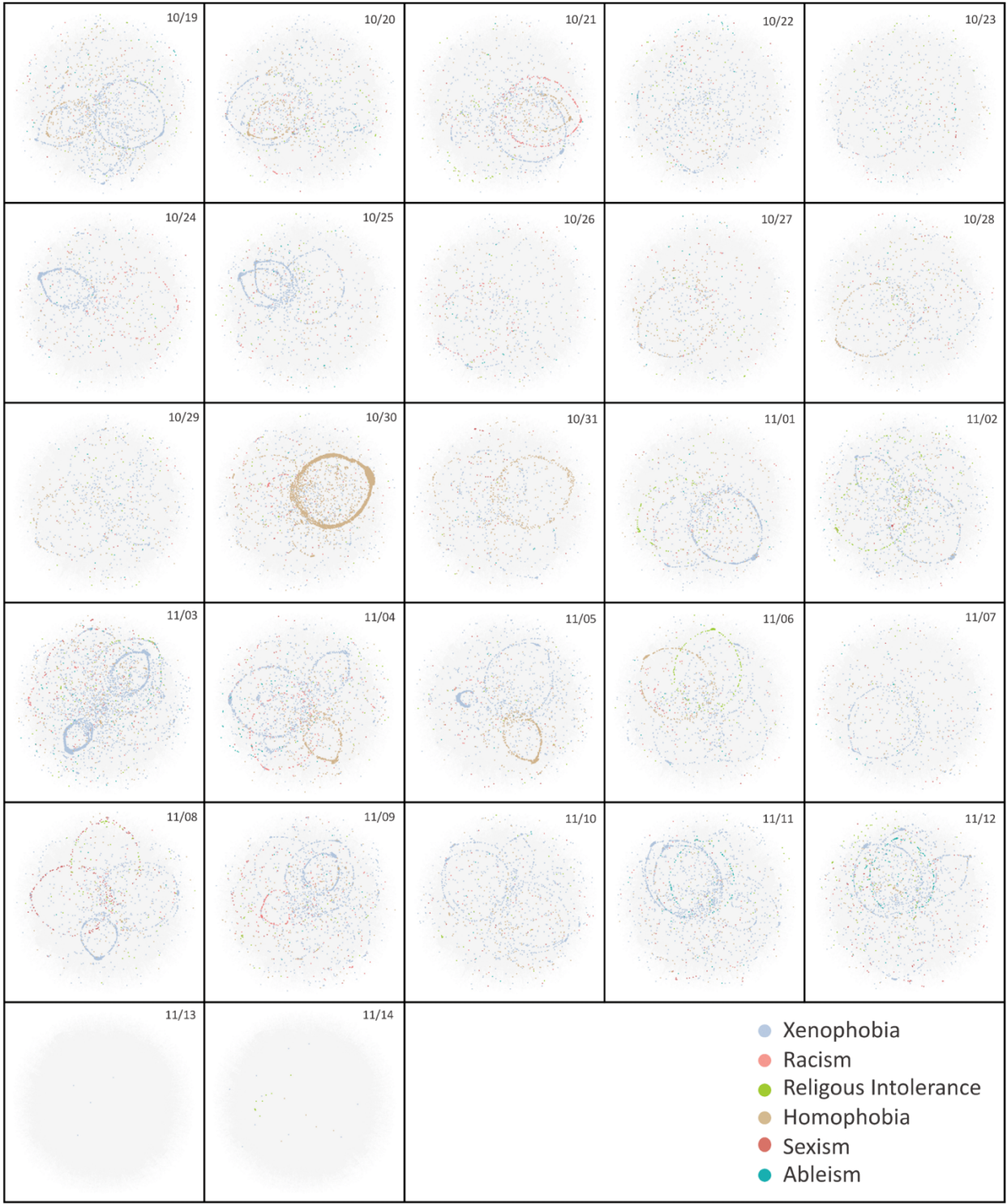

Figure 13. Seattle in the 2020 presidential race

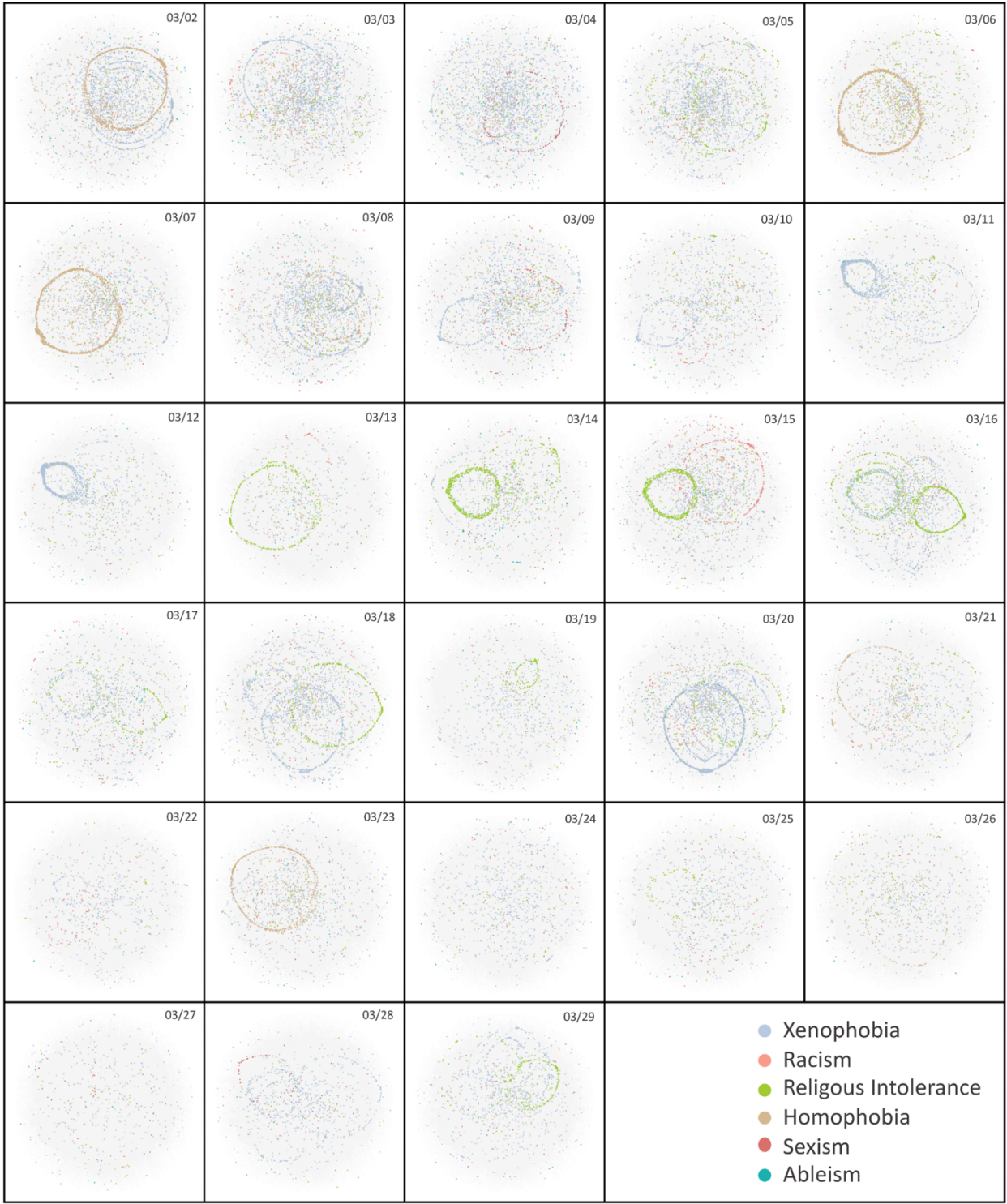

*Figure 14. San Francisco during the Coronavirus pandemic*